\documentclass[%
 reprint,
 amsmath,amssymb,
 aps,
]{revtex4-1}

\usepackage{graphicx}
\usepackage{dcolumn}
\usepackage{bm}
\usepackage{xspace}
\usepackage{multirow}
\usepackage{booktabs}
\usepackage{xcolor}
\usepackage{xfrac}
\usepackage[caption=false]{subfig}
\graphicspath{{./plots/}}

\usepackage{hyperref}
\hypersetup{
    colorlinks=true,
    linkcolor=blue,
    filecolor=blue,      
    urlcolor=blue,
    pdftitle={Overleaf Example},
    pdfpagemode=FullScreen,
    }

\usepackage{units}
\usepackage{comment}

\def\epos{\textsc{EPOS-LHC}\,\xspace}

\def\PalphaTtail{$P^{T}_{\rm tail}$\xspace}
\def\PalphaTtail{$P^{\alpha, T}_{\rm tail}$\xspace}

\newcommand{\x}[1]{%
  {}$
  \kern-2\mathsurround 
  $
  \binoppenalty10000 \relpenalty10000 #1
  {}$
  \kern-2\mathsurround 
  $
}

\begin{document}


\title{Time-Structured Tail Probabilities for Ultra-High-Energy Gamma–Hadron Discrimination in Water-Cherenkov Arrays}

\author{Ruben Concei\c{c}\~{a}o}
\address{Departamento de F\'isica, Instituto Superior T\'ecnico (IST), Universidade de Lisboa, Av.\ Rovisco Pais 1, 1049-001 Lisbon, Portugal}
\address{Laborat\'{o}rio de Instrumenta\c{c}\~{a}o e F\'{i}sica Experimental de Part\'{i}culas (LIP), Av.\ Professor\ Gama Pinto 2, 1649-003 Lisbon, Portugal}

\author{Pedro J. Costa}
\email{pmcosta@lip.pt}
\address{Departamento de F\'isica, Instituto Superior T\'ecnico (IST), Universidade de Lisboa, Av.\ Rovisco Pais 1, 1049-001 Lisbon, Portugal}
\address{Laborat\'{o}rio de Instrumenta\c{c}\~{a}o e F\'{i}sica Experimental de Part\'{i}culas (LIP), Av.\ Professor\ Gama Pinto 2, 1649-003 Lisbon, Portugal}

\author{M\'ario Pimenta}
\address{Departamento de F\'isica, Instituto Superior T\'ecnico (IST), Universidade de Lisboa, Av.\ Rovisco Pais 1, 1049-001 Lisbon, Portugal}
\address{Laborat\'{o}rio de Instrumenta\c{c}\~{a}o e F\'{i}sica Experimental de Part\'{i}culas (LIP), Av.\ Professor\ Gama Pinto 2, 1649-003 Lisbon, Portugal}

\date{\today}

\begin{abstract}
Gamma--hadron discrimination based on shower observables is essential for identifying gamma-ray astrophysical sources at the highest energies. In this work, we introduce \PalphaTtail, a new discrimination variable for ultra-high-energy photon searches within the framework of a water-Cherenkov detector (WCD) array. The observable extends signal-integrated methods by incorporating the time structure of WCD traces, using cumulative signal distributions.

Using simulated proton- and gamma-induced air showers at energies around \(10^{17}\,\mathrm{eV}\), we evaluate the performance of \PalphaTtail and compare it with established WCD-based observables such as \(S_b\), risetime-based variables, and the SWGO-inspired, \(P^\alpha_{\rm tail}\). The new variable attains a background contamination of roughly \( 2\times10^{-2}\) at 50\% gamma efficiency, improving upon existing WCD-only methods by nearly a factor of five and approaching the performance of an idealized muon-isolating reference. These results demonstrate the effectiveness of exploiting time-resolved signal tails to enhance ultra-high-energy photon searches in sparse surface arrays.

\end{abstract}

\pacs{Valid PACS appear here}
\maketitle


\section{Introduction}
\label{sec:intro}
The detection of gamma rays above $\sim 10^{16}\,$eV offers a powerful avenue for probing the most energetic astrophysical environments and for testing fundamental particle interactions at energies beyond the reach of laboratory accelerators. Photon-induced air showers provide insights into the acceleration mechanisms of cosmic-ray sources, the characteristics of intergalactic propagation, and the expected flux of secondary particles from processes such as the GZK effect. Despite their relevance, no multi-PeV photon has been unambiguously identified to date, and current limits rely critically on the ability to discriminate gamma-ray–induced extensive air showers (EAS) from the overwhelmingly more abundant hadron-induced background ~\cite{AugerPhotonReview}.

At energies above $\sim10^{16}\,$eV, photon-induced EAS differ from hadronic-induced showers in several key aspects: they reach their maximum development deeper in the atmosphere, contain significantly fewer muons, and produce narrower lateral and temporal particle distributions at ground level. Surface detector arrays, particularly those composed of Water-Cherenkov Detectors (WCDs), are sensitive to these features through both the spatial and temporal structure of the recorded signals. This has motivated the development of observables designed to exploit water-Cherenkov detector information for gamma--hadron separation, especially in experiments where fluorescence or hybrid information is not always available.

Within the Pierre Auger Observatory~\cite{PierreAuger:2015eyc}, variables such as the risetime-to-distance slope (ToD)~\cite{AveragedRiseTime} and the signal-based $S_b$ parameter~\cite{Auger_Sb} have become standard tools for composition and photon searches using WCD-only data. More recently, \textit{tail-probability} techniques have been introduced in the context of the Southern Wide-field Gamma-ray Observatory (SWGO)~\cite{SWGO:2025taj}, where the $P^\alpha_{\rm tail}$ variable leverages the presence of large, localized signals in detectors far from the shower core, typically produced by energetic muons or electromagnetic sub-showers, to enhance the separation between gamma- and hadron-induced events.

However, these methods rely primarily on integrated station signals and therefore do not fully exploit the rich time structure encoded in WCD traces. Temporal information is particularly valuable in sparse arrays such as the Pierre Auger Observatory, where the number of triggered stations is limited and maximizing the information extracted from each detector becomes essential. Late-arriving muons and high-energy sub-showers leave characteristic imprints in the time evolution of station signals, offering an additional and largely untapped discriminator between electromagnetic and hadronic primaries.

Gamma-induced showers, characterized by a lower transverse momentum and reduced muon content, tend to produce more spatially compact and temporally concentrated signals in the WCDs. In contrast, hadronic showers generate a broader temporal distribution of signals due to a larger number of muons arriving over an extended time window. Furthermore, sub-showers initiated by high-energy muons can produce late, high-amplitude signals that are more prevalent in hadron-induced shower events.
By incorporating this temporal information into tail-probability methods, it is possible to enhance the sensitivity to the muonic component of hadronic showers, thereby improving gamma--hadron discrimination.

In this work, we introduce a new observable, \PalphaTtail, designed to incorporate time-resolved information into tail-probability methods. The variable is constructed from cumulative signal distributions defined in radial distance from the shower core and in discrete time bins along the WCD trace. This approach enhances sensitivity to temporally extended, high-amplitude signals predominantly associated with the muonic component of hadronic showers. We evaluate the performance of \PalphaTtail using simulated proton- and gamma-induced air showers at energies around \(10^{17}\,\mathrm{eV}\), reconstructed with the Auger Offline framework under realistic detector conditions, and compare it to established gamma--hadron discrimination observables in the literature.

The structure of the paper is as follows: section~\ref{sec:simulations} describes the simulation set, event reconstruction, and detector modeling; section~\ref{sec:Ptail} introduces the \PalphaTtail variable and details the selection criteria applied to ensure signal quality; in section~\ref{sec:PTtailResults} it is presented the discrimination performance and comparison with standard observables; finally, conclusions and prospects for future applications are given in Section~\ref{sec:conclusions}.


\section{Simulation and reconstruction framework}
\label{sec:simulations}

CORSIKA (version 7.5600)~\cite{CORSIKA} was used to simulate gamma– and proton-induced extensive air showers with zenith angles around $\theta \sim 30^\circ$, assuming an observation level at $1400\,$m a.s.l. (approximately the Pierre Auger Observatory~\cite{PierreAuger:2015eyc} altitude).  Hadronic interactions were modeled with \textsc{UrQMD}~\cite{Bleicher:1999xi} at low energies and \epos~\cite{2015_Pierog_eposlhc} at high energies. Given the involved energies, the showers were generated using a thinning algorithm of \(10^{-6}\).

The Pierre Auger Observatory's surface detector array was simulated using the Auger Offline framework (version 0d68f71 2022-07-29) \cite{Offline_Framework}. Only the densest part of the array, with $433\,{\rm m}$ spacing between stations, was taken into account. This array, commonly referred to as the SD-433 array, covers an approximately circular region of $\sim  866\,{\rm m}$ radius with a uniform  $6.27\times 10^{-5}$ fill factor (FF), defined as the fraction of the shower collection area covered by stations.

The simulated and reconstructed zenith and energy ranges were chosen according to the angular and energy resolution of the SD-433 surface array of the Pierre Auger Observatory. The angular resolution, $\sigma_\theta$, varies from $1.8^\circ$ at $10^{16}\,$eV to $0.5^\circ$ at $10^{17}\,$eV~\cite{SD_433_Ang_Res}, while the energy resolution, $\sigma_E$, is of the order of $12\%$ around $10^{17}\,$eV~\cite{SD_433_E_Res}.

Proton showers were simulated with energies uniformly distributed over a range with a width of $2\sigma_E$ around $10^{17}\,$eV, corresponding to $10^{16.92}$–$10^{17.07}\,$eV. For gamma rays, a shifted interval of $10^{17.02}$–$10^{17.17}\,$eV was used to compensate for the known bias in their energy reconstruction ~\cite{PierreAuger:2025jwt}. Only events with reconstructed energies in the interval $[10^{16.97},\,10^{17.03}]$\,eV were retained for the analysis.

Zenith angles were sampled uniformly in the range $28.5^\circ$–$31.5^\circ$, and events were required to have reconstructed zenith values within $[29^\circ,\,31^\circ]$. Azimuth angles were drawn uniformly from $-180^\circ$ to $180^\circ$.

\begin{figure}[!t]
\centering
\includegraphics[width=\linewidth]{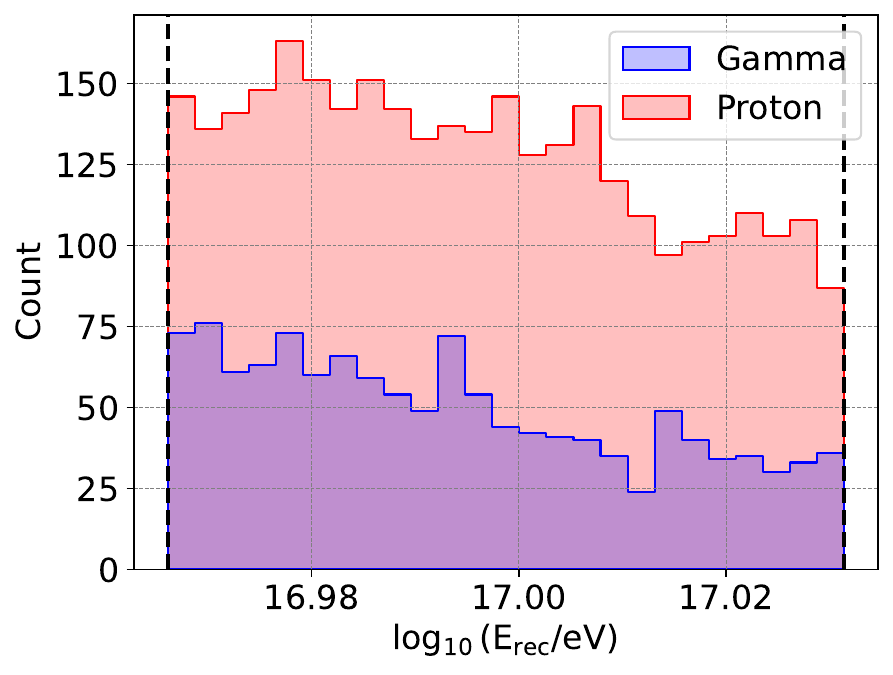}
\caption{\label{fig:rec_E_dists} Reconstructed energy distributions of the set of simulated proton (red) and gamma-ray (blue) events.
}
\end{figure}

\begin{figure}[!t]
\centering
\includegraphics[width=\linewidth]{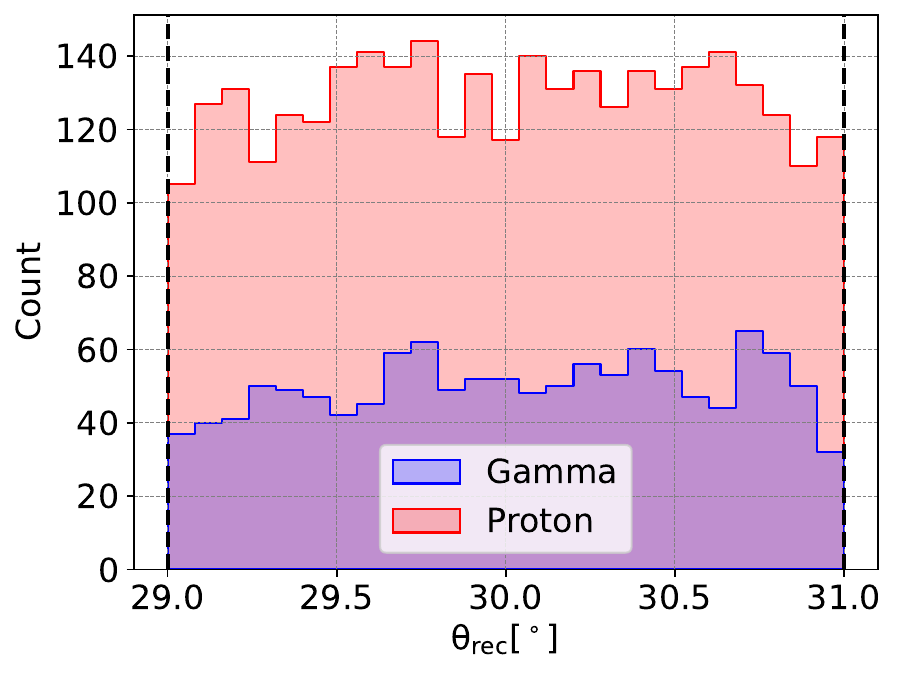}
\caption{\label{fig:rec_theta_dists} Reconstructed zenith distributions of the set of simulated proton (red) and gamma-ray (blue) events. 
}
\end{figure}

After all selection criteria, the dataset consists of 3211 proton-induced and 1243 gamma-induced showers. The distributions of reconstructed energy and reconstructed zenith for these events are shown in Fig.~\ref{fig:rec_E_dists} and Fig.~\ref{fig:rec_theta_dists}, respectively.


\section{The discriminant variable: \PalphaTtail}
\label{sec:Ptail}

Recently, a high-performance gamma--hadron discrimination and mass-composition observable, \(P^{\alpha}_{\rm tail}\)~\cite{Ptail}, was introduced in the framework of the Southern Wide-field Gamma-ray Observatory (SWGO)~\cite{SWGO:2025taj}. This variable builds upon a method originally developed by the IceTop/IceCube Collaboration~\cite{icetop}, motivated by the observation that, for events with similar reconstructed energies, WCD stations located far from the shower core typically register stronger signals when struck by energetic subshowers. Dominated by muons and high-energy electromagnetic particles, these subshowers are more abundant in hadron-induced air showers and therefore provide a discriminating handle~\cite{HAWC:2022hny}.

The \(P^{\alpha}_{\rm tail}\) observable is defined as
\begin{equation}
    P_{\rm tail}^{\alpha} \equiv \sum_{i=1}^{n} \left(P_{{\rm tail},i}\right)^{\alpha},
    \label{eq:PtailSWGO}
\end{equation}

where \(n\) is the number of stations with signal and \(P_{{\rm tail},i}\) is one minus the probability that the signal measured in the \(i\)-th WCD station falls within the upper tail of the reference signal distribution. The parameter \(\alpha\) controls the weight assigned to stations with large tail probabilities, i.e.\ those for which \(P_{{\rm tail},i}\!\to\!1\)~\cite{P_alpha}. For \(\alpha=1\), the observable reduces to the sum of probabilities. 

The quantity \(P_{{\rm tail},i}\) is computed, for each event and each station \(i\), as
\begin{equation}
    P_{{\rm tail},i} = C_{r_i}(S_i),
    \label{eq:Ptaili}
\end{equation}
where \(S_i\) is the signal recorded in the \(i\)-th station of the event. The function \(C_{r_i}\) denotes the normalized cumulative distribution of signals measured within a circular ring in the shower’s transverse plane, between a distance of \(r_i\) and \(r_i + 10\,\mathrm{m}\) from the shower axis. 
A cumulative distribution, \(C_{r_i}\), is computed for each radial bin, \(r_i\), and these are constructed from sets of hadronic shower events with similar reconstructed energies, allowing for a data-driven approach by using real data, should it be available.
Using simulated data for SWGO conditions~\cite{SWGO:2025taj}, it was found that gamma events exhibit smaller values of \(P^\alpha_{\rm tail}\) compared to hadronic events of similar reconstructed energy. 

The \(P^\alpha_{\rm tail}\) observable was originally developed for an array with a fill factor of \(12.5\%\), targeting events with reconstructed energies between \(10\,\mathrm{TeV}\) and \(1\,\mathrm{PeV}\). In contrast, the Pierre Auger Observatory operates with a fill factor nearly three orders of magnitude smaller, resulting in substantially fewer active stations per event ($\mathcal{O}(10)$ for vertical events). Furthermore, the relevant energies are about two orders of magnitude higher, yielding a much larger muon content per station. These muons also arrive with increased temporal dispersion due to the larger spatial extent of multi-PeV air showers.

\begin{figure}[!t]
\centering
\includegraphics[width=\linewidth]{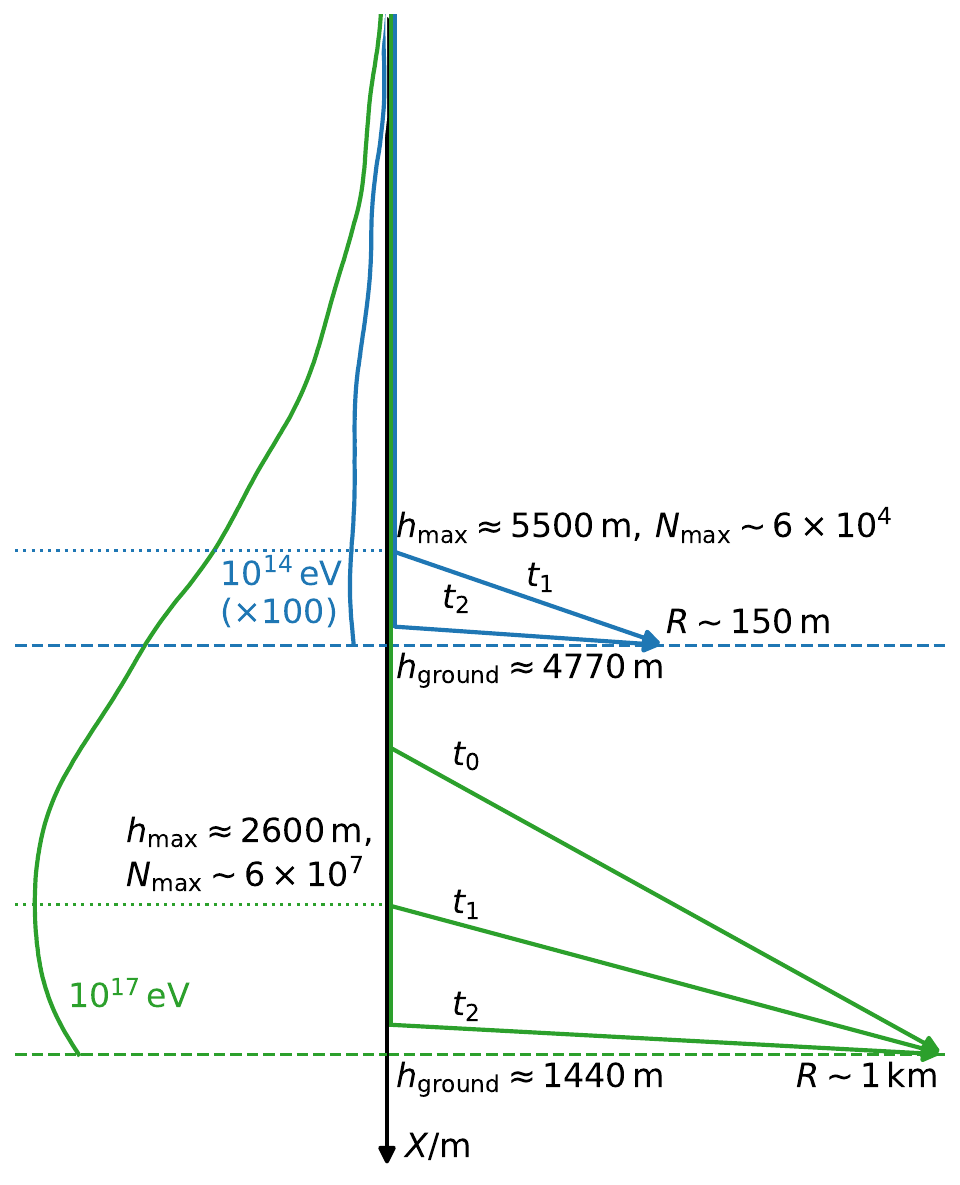}

\caption{\label{fig:EASTimeStruct} Diagram illustrating the differences in development and time structure of extensive air showers at different energies.
The left hand side roughly represents the typical longitudinal profile of the showers, while the right hand presents the characteristics of the showers at $X_{\rm max}$ (dotted lines), and at the ground level (dashed lines).
The difference in arrival time between two possible particle trajectories, $\Delta t \equiv t_2 - t_1$, is of the order of hundreds of nanoseconds for the lower energy shower ($100\,$TeV, blue) and of the order of a microsecond for the higher energy shower ($100\,$PeV, green).   
The number of particles in the longitudinal profile of the lower energy shower has been scaled up by a factor of 100 for visualization purposes.
}
\end{figure}

The difference in longitudinal development and time structure between extensive air showers with energies of $100\,\mathrm{TeV}$ and $100\,\mathrm{PeV}$ is illustrated in Fig.~\ref{fig:EASTimeStruct}. Showers with energies of the order of $100\,\mathrm{TeV}$ reach their maximum development higher in the atmosphere and produce a relatively small number of secondary particles that arrive at the observation level within a narrow spatial and temporal window. Their ground footprint typically extends over a few hundred meters in radius, with particle arrival times clustered within a few hundred nanoseconds.

In contrast, higher-energy showers develop deeper in the atmosphere and generate a substantially larger number of secondary particles, which reach the ground over both a wider spatial area and a broader time interval. Showers with energies around $100\,\mathrm{PeV}$ can cover ground areas with radii of order one kilometer, and particle arrival times can extend over intervals of up to several microseconds.

As illustrated in Fig.~\ref{fig:EASTimeStruct}, the maximum difference in arrival time between two particles, $\Delta t \equiv t_2 - t_1$, is typically $\mathcal{O}(400\,\mathrm{ns})$ for $100\,\mathrm{TeV}$ showers, whereas for $100\,\mathrm{PeV}$ showers it can reach values of order $2\,\mu\mathrm{s}$.

To account for these differences and fully leverage the time-resolved data available from WCDs, we introduce in this work a new observable, \PalphaTtail, which incorporates the time structure of the WCD traces recorded at each station. 

The \PalphaTtail observable is defined as
\begin{equation}
    P_{\rm tail}^{\alpha T} \equiv 
    \sum_{i=1}^{n_i} \sum_{j=1}^{n_j}
    \left(P_{{\rm tail},i,j}\right)^{\alpha},
    \label{eq:PtailT}
\end{equation}
where \(n_i\) is the number of active stations, \(n_j\) the number of time bins in the station trace, and \(P_{{\rm tail},i,j}\) denotes the one minus the probability that the signal measured in the \(j\)-th time bin of the \(i\)-th station falls within the upper tail of the corresponding reference signal distribution. As before, for \(\alpha = 1\), the observable reduces to the sum of tail probabilities over all stations and all trace bins.

The quantity \(P_{{\rm tail},i,j}\) is computed, for each event and each station \(i\), as
\begin{equation}
    P_{{\rm tail},i,j} = C_{r_i, t_j}(S_{i,j}),
    \label{eq:Ptailij}
\end{equation}
where \(S_{i,j}\) is the signal recorded in the \(j\)-th time bin of the trace of the \(i\)-th station. The function \(C_{r_i, t_j}\) denotes the normalized cumulative distribution of signals measured within a circular ring in the shower’s transverse plane between a distance of \(r_i\) and \(r_i + 50\,\mathrm{m}\) from the shower axis
and within a time interval beginning at \(t_j\) after the time of arrival of the shower-core and spanning \(25\,\mathrm{ns}\). The cumulative distributions for each radial–temporal bin can be constructed from sets of shower events with similar reconstructed energies, using real data when available or simulated data otherwise.

\begin{figure}[!htbp]
\centering
\subfloat[All cumulative distributions for time bins from $0$ to $1000\,\rm{ns}$.\label{fig:all_cumulatives_300}]{\includegraphics[width=0.5\textwidth]{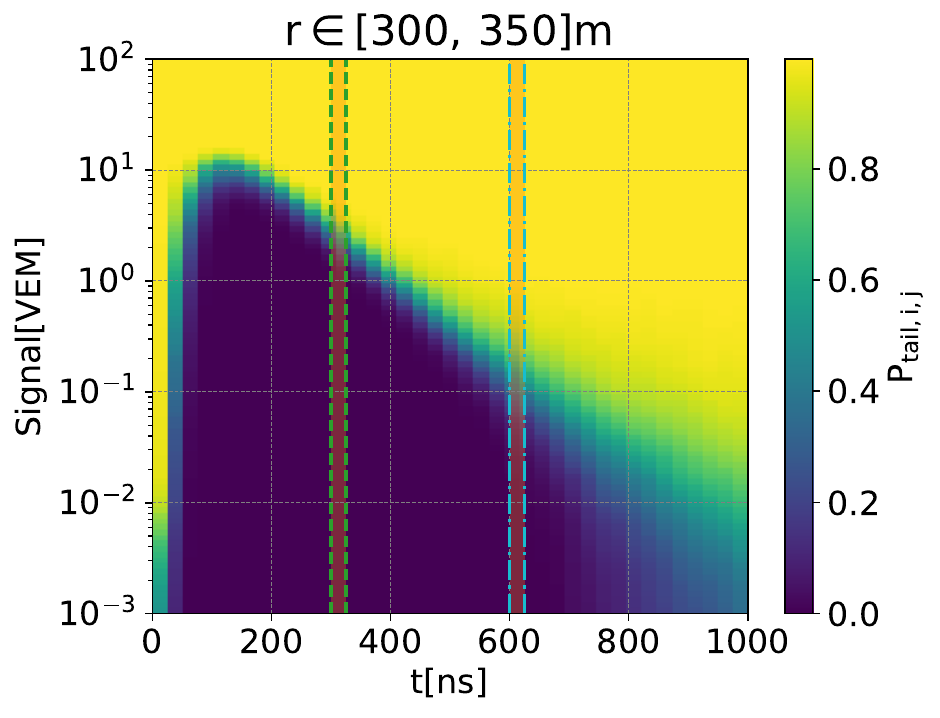}}\hfill

\subfloat[Selected individual cumulative distributions.\label{fig:select_cumulatives_300}] {\includegraphics[width=0.5\textwidth]{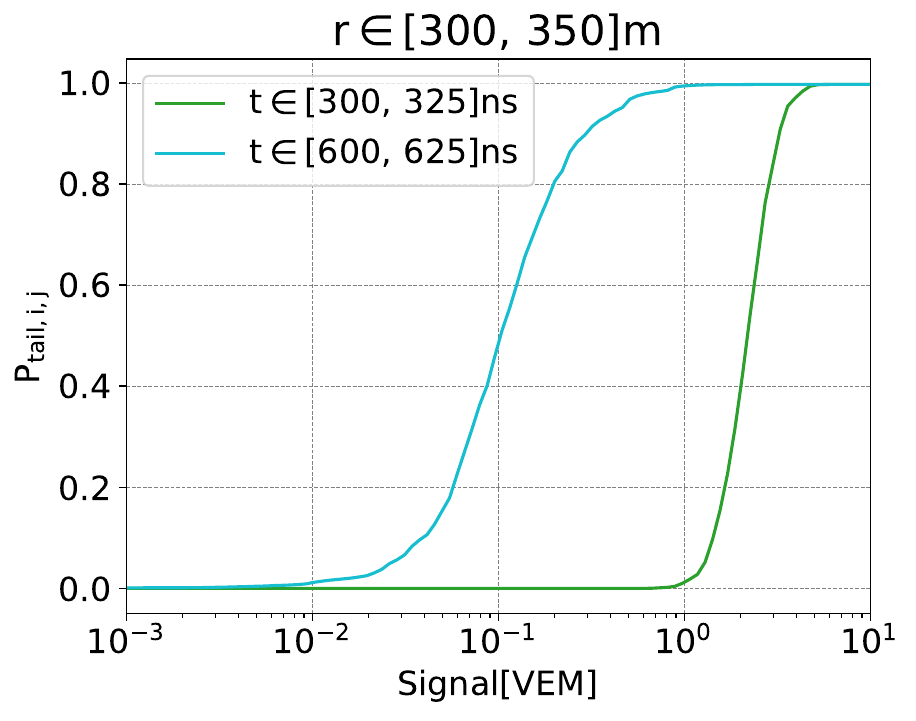}}
\caption{Cumulative signal distributions of the time trace for the ring spanning from $r_i=300$ m to $r_i=350$ m. The time of each bin is expressed in relation to the arrival time of the shower core. The two areas highlighted in the top figure correspond to the distributions shown in the bottom figure.
} \label{fig:cumulatives_300}
\end{figure}

\begin{figure}[!htbp]
\centering
\subfloat[All cumulative distributions for time bins from $0$ to $1000\,\rm{ns}$.\label{fig:all_cumulatives_500}]{\includegraphics[width=0.5\textwidth]{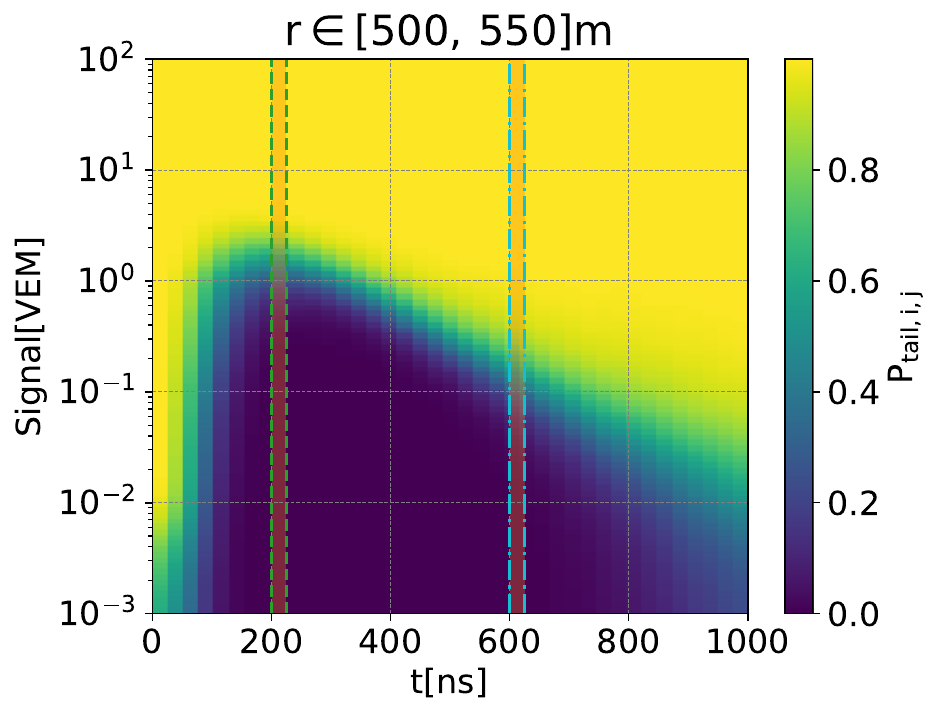}}\hfill

\subfloat[Selected individual cumulative distributions.\label{fig:select_cumulatives_500}] {\includegraphics[width=0.5\textwidth]{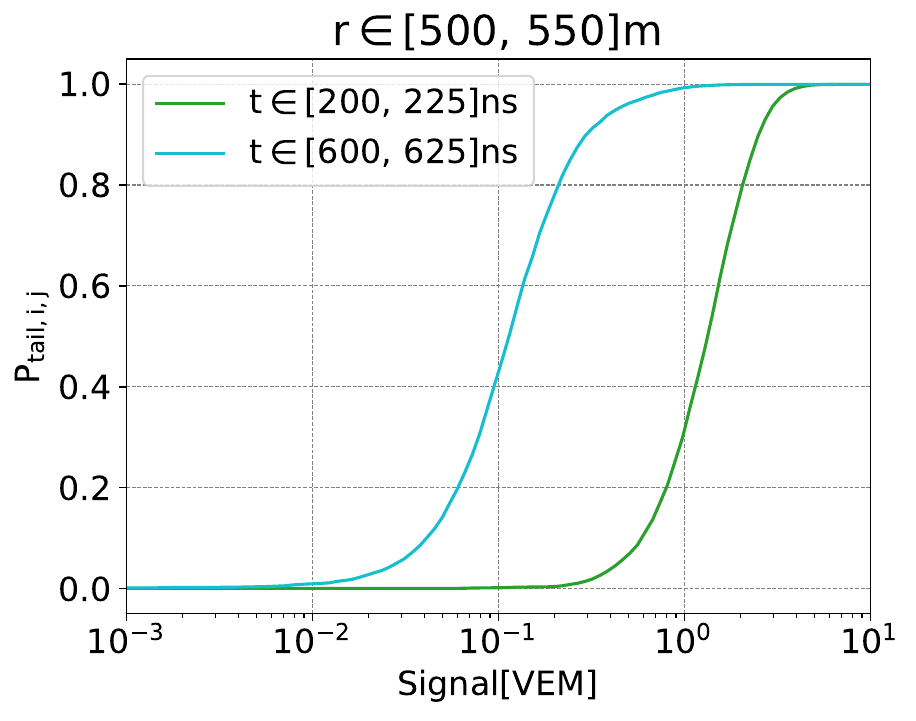}}
\caption{Cumulative signal distributions of the trace time bins for the ring spanning from $r_i=500$ m to $r_i=550$ m. The time of each bin is expressed in relation to the arrival time of the shower core. The two areas highlighted in the top figure correspond to the distributions shown in the bottom figure.
} \label{fig:cumulatives_500}
\end{figure}

As an illustration, Figs.~\ref{fig:cumulatives_300} and~\ref{fig:cumulatives_500} show the mean signal profiles and their corresponding normalized cumulative distributions for proton-induced showers in rings starting at \(r_i = 300\,\mathrm{m}\) and \(500\,\mathrm{m}\). These cumulative distributions correspond to the functions \(C_{r_i, t_j}\) entering Eq.~\eqref{eq:Ptailij}. It must be noted that the events used in the construction of these cumulatives correspond to the proton simulations described in section~\ref{sec:simulations}, without any of the quality cuts described below. 

To ensure that \PalphaTtail captures only high-quality and physically meaningful information, several selection cuts must be applied. These concern the distance of each station from the shower core, the time of the trace bin relative to the arrival of the core at the ground, the signal amplitude per bin, and the validity of the corresponding cumulative distributions.

\begin{figure}[!t]
    \centering
    \includegraphics[width=\linewidth]{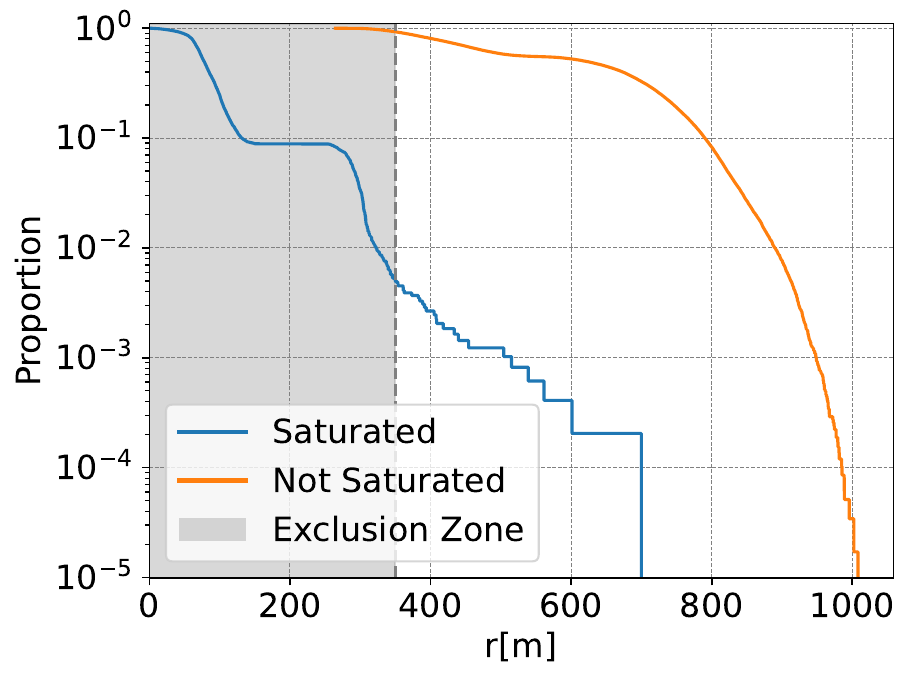}
    \caption{\label{fig:NbSaturatedStations} Reverse cumulative distributions of saturated (blue line) and non-saturated stations (orange line) for all events in the simulation set, as a function of the distance to the shower core along the shower plane. The shadowed area represents the region excluded by the cut in distance to the core.
    }
\end{figure}

A cut on the core distance is necessary to exclude the region where station saturation is likely to occur, while maintaining a consistent number of usable stations per event. That is, the number of stations considered in the computation of \PalphaTtail should not vary by more than a single station between the set of events being analyzed~\footnote{To mitigate the explicit dependence of the observable on the number of triggered stations, an alternative approach was explored in which $P_{\mathrm{tail},i,j}$ was parametrized as a function of the distance to the shower core and evaluated at a fixed reference distance, instead of being summed over stations and time bins. This procedure would provide a proxy for $P^{\alpha,T}_{\mathrm{tail}}$ that is independent of station multiplicity. However, owing to the small number of stations available in sparse arrays, the resulting fits were found to be insufficiently stable and did not yield competitive gamma--hadron discrimination.}. 
This avoids artificially enhancing \PalphaTtail simply because a few stations are saturated close to the core. Since the interpretation of saturated waveforms is ambiguous, such stations are excluded entirely. In this study, the minimum distance from the shower axis (in the shower plane) was set to \(350\,\mathrm{m}\). Given the \(50\,\mathrm{m}\) ring width used to construct the cumulatives, this choice preserves at least \(90\%\) of non-saturated stations while rejecting more than \(99\%\) of saturated ones. The cumulative radial distributions of saturated and non-saturated stations, which motivate this selection, are shown in Fig.~\ref{fig:NbSaturatedStations}.

A cut on the signal amplitude per trace bin is required to suppress contributions arising from baseline noise. Fluctuations in the baseline can otherwise lead to spurious nonzero probabilities in the construction of \PalphaTtail. To ensure that only physically meaningful signals are included, a minimum bin signal of \(0.4\,\mathrm{VEM}\) was adopted\footnote{One VEM corresponds to the signal recorded at the WCD by a vertically centered muon~\cite{PierreAuger:2005znw}.}.
This value was obtained maximizing the discrimination power of \PalphaTtail. It it important to note that this threshold is applied solely to the event-by-event computation of \PalphaTtail; it is not imposed when constructing the reference cumulative signal distributions, which must represent the full underlying distribution of trace amplitudes.

In addition, late-arriving and very inclined particles may occasionally produce large, localized spikes in the traces due to Cherenkov light striking the photomultiplier tubes directly, that is, without first reflecting off the tank walls. Although such spikes can originate from muons, they are more commonly associated with electrons and positrons interacting with the tank walls, and they can artificially enhance the values of \(P_{{\rm tail},i,j}\) for the affected stations. To avoid this contamination, only signals recorded within a restricted time window—between \(0\) and \(500\,\mathrm{ns}\) after the shower-core arrival time—are considered.

\begin{figure}[!t]
    \centering
    \includegraphics[width=\linewidth]{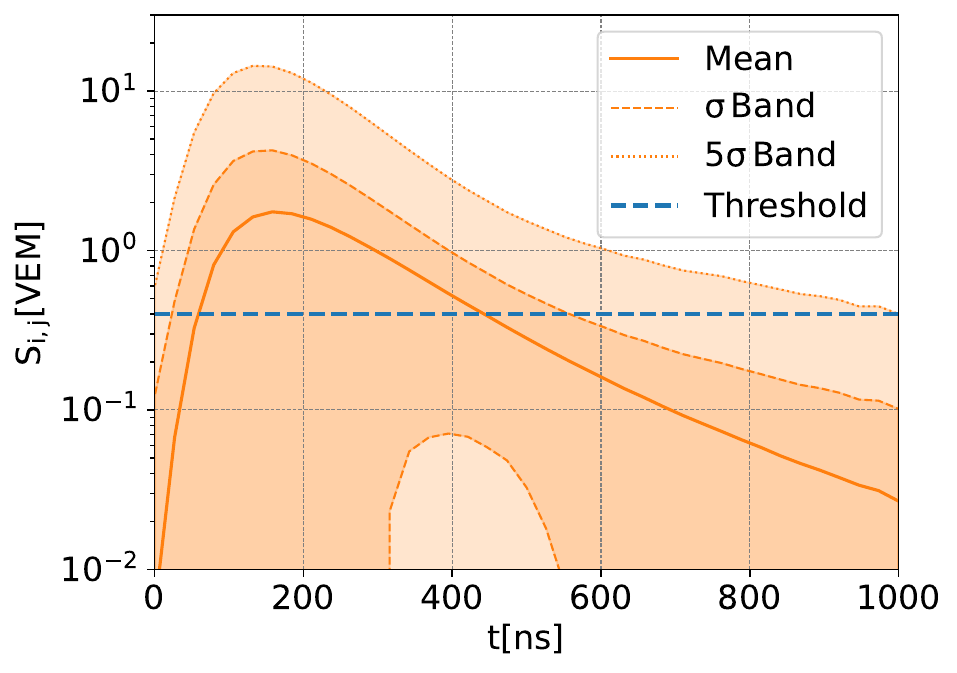}
    \caption{\label{fig:TraceBinAverageSignal} Average signal of non-saturated stations as a function of time after the shower-core arrival at the array. The blue dashed line indicates the signal threshold applied to suppress baseline fluctuations. The solid orange line shows the mean trace-bin signal, and the shaded band denotes the corresponding \(1\sigma\) and \(5\sigma\) fluctuations.
    }
\end{figure}

The choice of this time window is further supported by Fig.~\ref{fig:TraceBinAverageSignal}, which shows the average trace-bin signal as a function of time for all non-saturated stations in the simulation set. Beyond approximately \(500\,\mathrm{ns}\), the mean signal returns to a level consistent with baseline fluctuations, indicating that bins at later times are dominated by noise rather than by shower particles.


Finally, a cut must be applied to the cumulative distributions used in the computation of \PalphaTtail in order to suppress the overwhelming electromagnetic background present near the shower core shortly after the event reaches the array. Since \PalphaTtail is designed to enhance the contribution of muons by selecting signals in the upper tails of the trace-bin distributions, its performance degrades in regions where the muonic component is obscured by large electromagnetic fluctuations. In such cases, the values of \(P{{\rm tail},i,j}\) become dominated by the electromagnetic background rather than by the muon content.

To avoid this, only cumulatives satisfying
\begin{equation}
    C_{r_i, t_j}(S_{i,j} = 1\,\mathrm{VEM}) \geq 0.5
\end{equation}

are retained, where \(i\) labels the radial ring, \(j\) the trace-bin index, and \(S_{i,j}\) the signal recorded in the \(j\)-th time bin of the station in ring \(i\). This requirement ensures that the expected signal from a vertical muon is sufficiently prominent relative to the electromagnetic contribution in the corresponding radial–temporal bin. 
In quantitative terms, this cut guarantees that at least half of the signals in the cumulative distributions do not exceed \(1\,\mathrm{VEM}\), a threshold typically associated with muonic signals. 

\begin{figure}[!t]
    \centering
    \includegraphics[width=\linewidth]{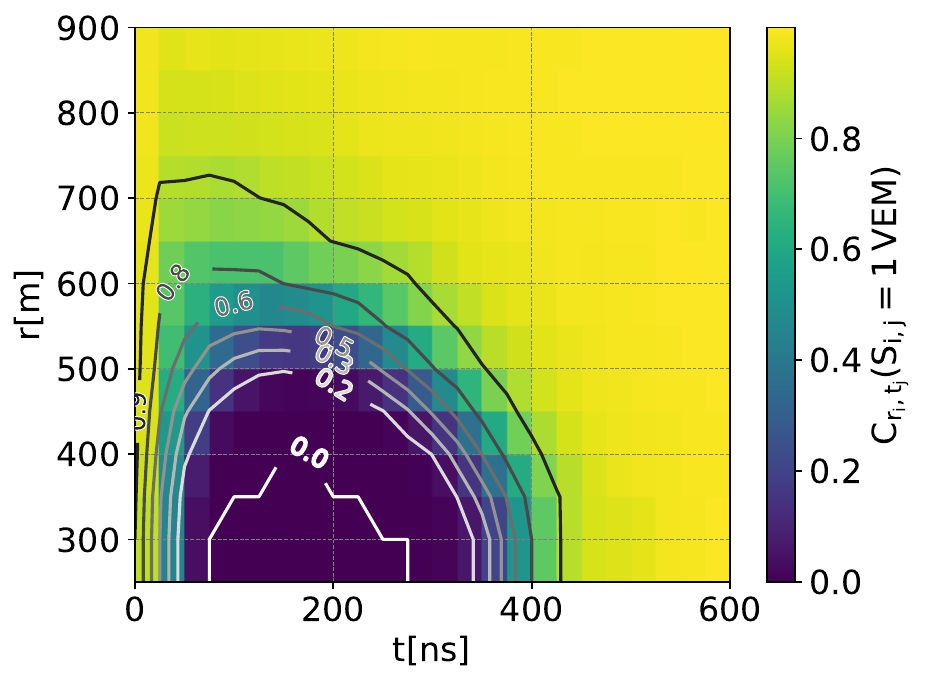}
    \caption{\label{fig:CumulativesAt1VEM} Values of the cumulative distributions evaluated at a trace-bin signal of \(1\,\mathrm{VEM}\). In the dark-shaded region, the expected signal from vertical muons is dominated by electromagnetic contributions.
    }
\end{figure}

The effectiveness of this selection is illustrated in Fig.~\ref{fig:CumulativesAt1VEM}, which shows the values of the cumulative distributions evaluated at \(1\,\mathrm{VEM}\) for all rings and time bins.

\section{Results on the discrimination power of \PalphaTtail}
\label{sec:PTtailResults}

The computation of \PalphaTtail, outlined in Sec.~\ref{sec:Ptail}, is performed on the simulation set described in Sec.~\ref{sec:simulations}. Surviving events are weighted according to their assumed fluxes, taken to follow power-law spectra of \(E_{\rm MC}^{-2.7}\) for protons and \(E_{\rm MC}^{-2}\) for gamma rays.

Additionally, to understand the gain in discrimination power of \PalphaTtail relative to other standard discrimination quantities relying exclusively on WCD data, we compute also the following quantities: risetime over distance ($\rm \overline{ToD}$), the quantity usually referred to as $S_b$, and \(P_{\rm tail}\).

The variable \(\overline{\mathrm{ToD}}\) (or \(\langle t_{1/2}/r \rangle\))~\cite{AveragedRiseTime} is a composition-sensitive observable defined as
\begin{equation}
    \overline{\mathrm{ToD}} = \frac{1}{N} \sum_{i=1}^{N} \frac{t_{1/2,i}}{r_i},
    \label{eq:RiseTimeOverDistance}
\end{equation}
where \(t_{1/2,i}\) is the risetime of station \(i\)—the time required for the integrated signal to increase from \(10\%\) to \(50\%\)—and \(r_i\) is the distance of that station to the shower axis in the shower plane.
The observable \(\overline{\mathrm{ToD}}\) captures the temporal spread of signals in the WCDs, which tends to be larger for hadron-induced showers in relation to gamma showers, as the former have a higher muon content and consequently a more spread-out shower time structure. 

Another commonly used WCD-based observable is \(S_b\)~\cite{Auger_Sb}, which is the current standard within the Pierre Auger Collaboration for gamma--hadron separation when relying exclusively on WCD data, without the use of Machine Learning algorithms. It is defined as
\begin{equation}
    S_b = \sum_i S_i \left( \frac{r_i}{1000\,\mathrm{m}} \right)^b,
    \label{eq:Sb_def}
\end{equation}
where \(S_i\) is the signal recorded in the \(i\)-th station, \(r_i\) its distance to the shower axis, and \(b\) a free parameter optimized for discrimination power, typically \(b \simeq 4\)~\cite{Auger_Sb}. This observable exploits differences in the lateral signal distribution between hadron- and gamma-induced air showers, which arise primarily from their distinct muon content and overall shower development.

\begin{figure}[!t]
\centering
\includegraphics[width=1.\linewidth]{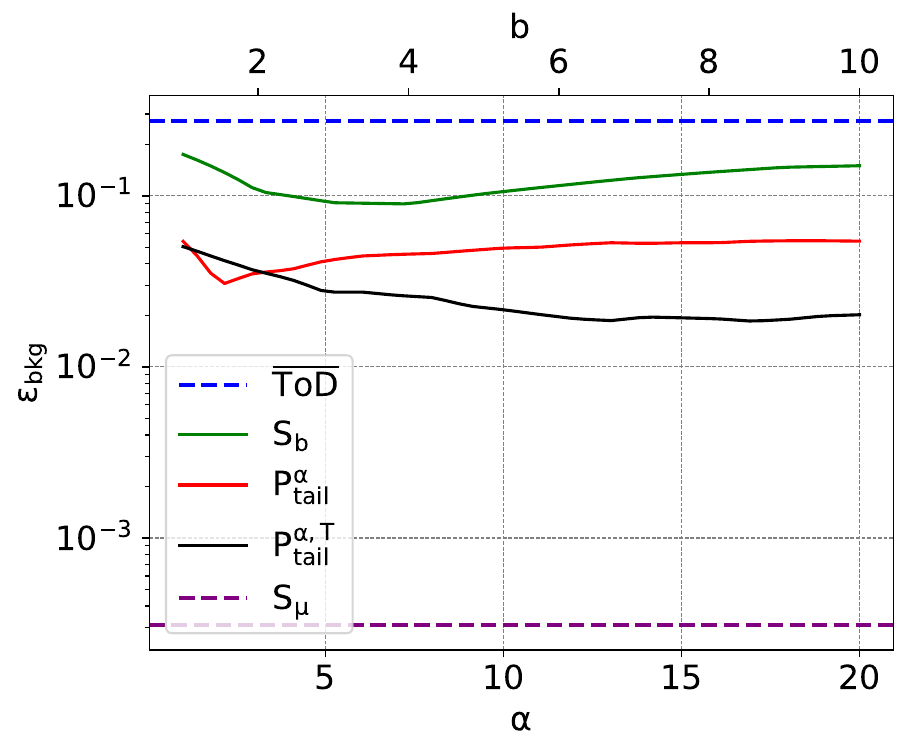}
\caption{\label{fig:bkg_rejection_efficiency} Background rejection of WCD-based gamma-hadron discrimination variables at $50\%$ signal efficiency (see legend for details, and text for the definitions of the discriminating observables).
}
\end{figure}

The gamma--hadron discrimination performance of all variables discussed above, including the SWGO observable \(P^\alpha_{\rm tail}\), is summarized in Fig.~\ref{fig:bkg_rejection_efficiency}, which shows the background contamination at \(50\%\) gamma efficiency. Since both the new observable and \(S_b\) depend on tunable parameters—\(\alpha\) for \PalphaTtail and \(b\) for \(S_b\)—their discrimination power is shown as a function of these parameters. For reference, \(S_\mu\), a curve corresponding to the discrimination power obtained when using the sum of the muonic signal in each station is also displayed, illustrating the behavior of an ideal muon detector.

According to Fig.~\ref{fig:bkg_rejection_efficiency}, the SWGO variable \(P^\alpha_{\rm tail}\) achieves a background contamination of approximately \(3\times 10^{-2}\) at \(\alpha=2\), while \(S_b\), reaches its optimal performance of \(\varepsilon_{\rm bkg} \approx 9\times 10^{-2}\) at \(b=4\). The observable introduced in this work, \PalphaTtail, outperforms both, attaining a background contamination of \(\varepsilon \approx 2\times 10^{-2}\) for \(\alpha \gtrsim 10\). Although \(P^{\alpha T}_{\rm tail}\) shows clear improvement over existing WCD-only observables, its performance naturally remains below that of the idealized reference observable \(S_\mu\), which is constructed directly from the simulated muonic component and is therefore unaffected by detector or electromagnetic fluctuations. At 50\% gamma efficiency, \(S_\mu\) rejects all proton events in the simulation set, corresponding to a discrimination level of \(\lesssim 3\times 10^{-4}\). This benchmark serves mainly to illustrate the upper limit attainable by a perfect muon-sensitive detector rather than representing a realistic target for WCD-only methods.

\begin{figure}[!t]
\centering
\includegraphics[width=\linewidth]{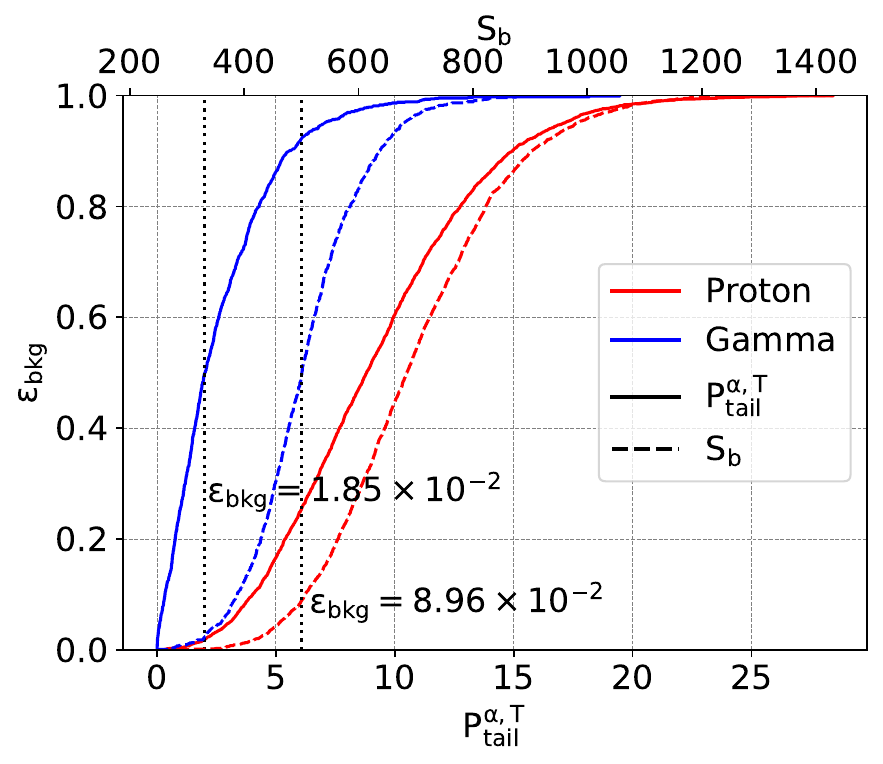}
\caption{\label{fig:cumulatives_Sb_and_PtailT} Cumulative distributions of the values of $S_b$ with $b=4$ (dashed lines), and of $P_{tail}^{\alpha T}$ with $\rm\alpha=17$ (solid lines). 
Gamma-rays are represented by blue lines and protons by red lines. The dotted lines denote the value at which $50\%$ efficiency in gamma-rays is reached. 
}
\end{figure}

The cumulative distributions corresponding to the optimal values of \(S_b\) and \PalphaTtail—namely \(b = 4\) and \(\alpha = 17\)—are shown in Fig.~\ref{fig:cumulatives_Sb_and_PtailT}. They illustrate the enhanced discrimination power of the latter observable, which outperforms the standard WCD-based discriminator by nearly a factor of five.

\section{Discussion and Conclusions}
\label{sec:conclusions}

In this work, we introduced \PalphaTtail, a new gamma--hadron discrimination observable designed to exploit the time structure of Water--Cherenkov Detector (WCD) traces in sparse surface arrays. By extending tail-probability techniques to two dimensions—radial distance and time trace—the method enhances sensitivity to late, high-amplitude energy deposits typically produced by muons and energetic sub-showers in hadron-induced air showers. 

Applied to realistic simulations of the SD-433 array of the Pierre Auger Observatory, and after standard reconstruction-quality cuts, \PalphaTtail exhibits a significant improvement over existing WCD-only discriminators. For an optimized choice of \(\alpha\), the observable achieves a background contamination of \(\varepsilon_{\rm bkg} \approx 2\times10^{-2}\) at \(50\%\) gamma efficiency—nearly a factor of five better than the Auger reference variable \(S_b\), and outperforming the SWGO-inspired \(P^\alpha_{\rm tail}\). These results demonstrate that incorporating time-resolved information provides a robust additional handle for gamma--hadron separation, even in an array where only \(\mathcal{O}(10)\) stations contribute to each event.

The comparison with the idealized benchmark observable \(S_\mu\) highlights both the progress and the remaining limitations. While \PalphaTtail moves significantly closer to muon-based discrimination than conventional WCD observables, it remains more than an order of magnitude less powerful than a detector capable of isolating the muonic component directly. This gap quantifies the irreducible influence of electromagnetic fluctuations on WCD signals and sets a realistic target on the performance achievable without dedicated muon detectors. It also underscores the potential of combining timing, signal-shape, and spatial-correlation information to further approach the ideal limit.

Several aspects merit further investigation. First, the present study focuses on near-vertical events (\(\theta \approx 30^\circ\)); more inclined showers, with their enhanced muon dominance and broader temporal profiles, may offer even greater discrimination potential for time-structured observables. 

Second, the construction of the cumulative distributions—and therefore the definition of \PalphaTtail—depends on the hadronic interaction models used in simulation. However, given the large statistics of cosmic-ray data and the, up to now, absence of detected multi-PeV gamma-ray events~\cite{PierreAuger:2025jwt}, these cumulative distributions can in practice be derived directly from data. In this case, only the determination of the optimal value of \(\alpha\) would necessarily rely on simulation studies, or alternatively on dedicated calibration campaigns aimed at quantifying the level of electromagnetic contamination in WCD signals. Such studies could be performed using the scintillator surface detectors deployed in the AugerPrime upgrade~\cite{AugerPrime}, or the MARTA prototype~\cite{MARTA, MARTA_RPC}, which incorporates resistive plate chambers beneath the WCD tank to directly measure the muonic component of air showers. Both detection systems are already commissioned and could be employed by the Pierre Auger Collaboration in the near future to refine the calibration and optimization of the observable proposed in this work.

Finally, the fill factor of the detector plays a central role in determining both the number of available stations and the statistical stability of the cumulative distributions. Arrays with higher station density—such as the proposed Project for Extreme PeVatrons Searches (PEPS) experiment~\cite{PEPS,PEPS2} for gamma-ray detection around \(10\,\mathrm{PeV}\)—could further enhance the performance of the method by providing more stations per event and reducing fluctuations in the radial–temporal bins.

In summary, \PalphaTtail represents a substantial advance in WCD-based gamma--hadron discrimination, providing improved performance in multi-PeV photon searches using only surface detector data. By incorporating time-resolved information into tail-probability methods, the observable captures physical features associated with the muonic shower component that remain inaccessible to standard integrated-signal approaches. Its demonstrated performance in an Auger-like array establishes its viability for current detectors, while its conceptual framework naturally extends to future, denser surface arrays. These results highlight the scientific potential of time-domain observables in multi-PeV cosmic-ray physics and open new avenues for improving photon searches in large-scale WCD arrays.

\section*{Acknowledgments}
We thank the Pierre Auger Collaboration for valuable discussions and for providing access to their simulation and reconstruction software. We also thank Vladim\'ir Novotn\'y for a careful reading of the manuscript.
This work has been funded by Fundação para a Ciência e Tecnologia, Portugal, under project \url{https://doi.org/10.54499/2024.06879.CERN}.
P.~C. is grateful for the financial support by FCT under UI/BD/153576/2022.

\bibliography{references}

\end{document}